\documentclass[preprint]{amsart}
\usepackage{amsmath}
\usepackage{mathrsfs}
\usepackage{amssymb}
\usepackage{mathtools}

\newtheorem{defn}{Definition}

\newtheorem{thm}[defn]{Theorem}

\def\wtK{K}
\title{Instability of stationary closed strings 
winding around flat torus in five-dimensional Schwarzschild~spacetimes}
\author{Mitsuharu Hasegawa \and Daisuke Ida}
\address{Department of Physics, Gakushuin University, Tokyo 171-8588}
\date{\today}
\begin{document}
\maketitle
\markright{\textsc{Instability of stationary closed strings in 5 dimensional Schwarzschild spacetimes}}

  \begin{abstract}
Linear perturbations for a one parameter family of 
stationary,  closed Nambu-Goto strings 
winding around a flat torus in the five-dimensional Schwarzschild spacetime
have been studied. It has been shown that this problem is solvable
in the sense that frequency spectra and perturbation modes can be
expressed only with arithmetic operations and radicals.
It has been proven that the Nambu-Goto strings belonging to this family
are always unstable, no matter how they are located at almost flat region
distant from the event horizon.
  \end{abstract}

\section{Introduction}
One promising direction in which to get a better grasp of the effect of gravitation
is brought about by the investigation of test objects in curved spacetimes,
such as test particles, strings, or membranes.
Among these, the motion of a test particle is described by the
geodesic equations consisting of ordinary differential equations, which are easily accessible.
Accordingly, the analysis of the test particles has been worked out 
in various background gravitational fields.

A natural generalization of the geodesic equation is given by
the Nambu-Goto equation for test strings or membranes describing such
test objects~\cite{Got71,Nam74}.
While the motion of geodesic particles is described by the world line
with the extremal spacetime interval, the Nambu-Goto strings or membranes
are characterized by their extremal area or volume for world sheets.
In a unified point of view, these  geodesic particles, Nambu-Goto strings/membranes are regarded as
the harmonic mappings for isometric embeddings
from lower-dimensional spacetimes
into the spacetime. 

The Nambu-Goto strings or membranes are considered to arise naturally
in our Universe. They correspond to the thin wall approximation 
of the  topological defect
produced via the symmetry breaking of the gauge interactions in the standard model
of elementary particles.
Their analytic solutions  are less known 
since they are subject to  partial differential
equations. In cosmological applications, the main approach relies on numerical
simulations, which are very useful for getting insight into the scenario for
the structure formation in our Universe~\cite{VS95}.

It is known that considerable simplification occurs for the equation of motion
of Nambu-Goto strings when the isometry of the background spacetime
acts on the string world sheet, in which case the system reduces to that of
particle motion in a certain lower-dimensional manifold~\cite{CFH91,KF08,KKI15}.
In this direction, many interesting analytic solutions for the Nambu-Goto equation have been
found. In particular, stationary string solutions
in stationary black hole spacetimes are extensively studied~\cite{FSZH89,LS95,dVE96,FHDV97}. 
These are regarded as 
final equilibrium configurations of Nambu-Goto
strings in the presence of a black hole~\cite{OIKNS08}.
Initially dynamical strings would radiate their
energy due to some dissipative processes such as emission of Nambu-Goldstone bosons, which, however, are effects that are neglected in the test string approximations.  Namely,
some strings would
fall into the black hole and others,
via such dissipative processes, 
would settle down to final stable configurations, which would be described by
stationary solutions.
Hence, we are interested in the stability of these stationary string configurations in curved background spacetimes.

Since the Nambu-Goto equation reduces to the linear wave equation in flat backgrounds, its linear perturbation is also subject to the linear wave equation. So, any stationary strings in flat backgrounds would be stable under small fluctuations. In the presence of a black hole,
we could, however, not expect stability for stationary strings. Hence,
we have to check the stability of stationary solutions separately.

The linear perturbations of the Nambu-Goto strings are formulated by 
Guven~\cite{Guv93}, 
and the stability problem for Nambu-Goto strings has been analyzed 
by many authors~\cite{Lar95,LF94,KL06,BKP17}.
We also follow  Guven's approach in order to study the stability problem of
stationary strings.

The subject of this paper is the stability problem of
a closed string winding around a flat torus embedded in five-dimensional nonrotating
black hole background, which is a nontrivial solution of Nambu-Goto equations recently discussed by Ref.~\cite{BEI08}.
This string solution is characterized by a
parameter corresponding to the distance between
the string and the event horizon.
We show that this class of string solutions allows
fully symbolic treatment for linear perturbations and that it can be proven that
these configurations are
always unstable. This means that the presence of
a black hole may switch on the instability of  strings even if they are  localized in an almost
flat region distant from the event horizon,
which might be unexpected with naive considerations.

This paper is organized as follows: Section 2 reviews the linear perturbation formalism developed by Guven for self-contained-ness.
Section 3 studies the stability of corresponding closed string solutions in the five-dimensional flat spacetime as a reference.
Section 4, which includes our main result, 
investigates the closed string solutions
in five-dimensional Schwarzschild spacetime,
and gives a proof that they are all unstable.
Finally, Section 5 presents our conclusions.

\section{Review of linear perturbations of Nambu-Goto membranes}
First, we review the formulation of linear perturbation for Nambu-Goto
membranes in general settings developed by Guven~\cite{Guv93}.

Let $M$ be a $m$-dimensional spacetime, and let $g$ be a Lorentzian
metric on $M$ with the signature $(-,+,\dots,+)$.
We consider the isometric embedding of an $n$-dimensional differentiable
manifold $N$ 
into $M$, where $2\le n\le m-1$. In terms of the local coordinates,
this embedding would be expressed as
\begin{align*}
x^\mu=X^\mu(\xi^a)
\end{align*}
where $x^\mu$'s $(\mu=1,\dots,m)$ are local coordinates on $M$,
while $\xi^a$'s $(a=1,\dots,n)$ are those on $N$.
We consider only the timelike embeddings so that
the spacetime metric $g$ induces the Lorentzian metric
\begin{align}
G_{ab}=g_{\mu\nu}(D_aX^\mu)D_bX^\nu
\end{align}
on $N$,
where $D_a$ denotes the $G$ connection on $N$.

The Nambu-Goto membranes are defined as the embeddings those extremize the
action
\begin{align*}
S[X^\mu]=-T\int d^n\xi \sqrt{|G|},
\end{align*}
where $T$ is a positive constant that does not play any special roles in the
present argument, and $G$ is abbreviation for $\operatorname{det}G$.
This leads to the Euler-Lagrange equation,
\begin{align*}
D_aD^a X^\mu+\Gamma^\mu_{\nu\lambda}(D_aX^\nu)(D_bX^\lambda)G^{ab}=0,
\end{align*}
where $\Gamma^\mu_{\nu\lambda}$
denotes the restriction of the Christoffel symbol for the $g$ connection on $N$.
Note that $D_aX^\mu$ corresponds to the $x^\mu$ component of the coordinate basis
$\partial/\partial \xi^a$ of the tangent space of $N$.

The extrinsic curvature of the embedding $N\xhookrightarrow{} M$ is measured in terms of
the second fundamental form
\begin{align}
K^\mu{}_{ab}=D_a D_b X^\mu+\Gamma^\mu_{\nu\lambda}(D_aX^\nu)D_bX^\lambda,
\label{DDX}
\end{align}
defined on $N$.
This is a symmetric tensor field on $N$ i.e.
\begin{align*}
K^\mu{}_{ab}=K^\mu{}_{(ab)}
\end{align*}
holds, and 
this can also be seen as a normal vector of $N$ parametrized by $(a,b)$
in the sense that
\begin{align*}
g_{\mu\nu}K^\mu{}_{ab}D_cX^\nu=0
\end{align*}
holds on $N$, as is easily confirmed.
The Nambu-Goto equation just says that the trace of the extrinsic curvature
vectors is zero:
\begin{align*}
K^\mu{}_{ab}G^{ab}=0.
\end{align*}

Denote by $X^\mu$ the unperturbed solution to the Nambu-Goto equation.
Our purpose here is to write down the linearized Nambu-Goto equation for 
small deviation $\delta X^\mu$ from $X^\mu$.
Obviously, it is sufficient to assume that $\delta X^\mu$ should be normal
to $N$, since the tangential component of $\delta X^\mu$ with respect to $N$
corresponds to diffeomorphism on $N$, which is uninteresting.
Hence, we set
\begin{align*}
\delta X^\mu=f^A n_A{}^\mu,
\end{align*}
in terms of $m-n$ differentiable functions $f^A$ $(A=1,\dots,m-n)$ on $N$,
where $n_A{}^\mu$'s constitute an orthonormal frame field for the normal bundle on $N$, i.e. such that
\begin{align*}
g_{\mu\nu}n_A{}^\mu D_aX^\nu&=0,\\
g_{\mu\nu}n_A{}^\mu n_B{}^\nu&=\eta_{AB},\\
\eta_{AB}&=\operatorname{diag}(1,\dots,1).
\end{align*}
Note that there is $O_{m-n}$ gauge freedom choosing $n_A$'s so that
the resultant equation for $f^A$'s should be covariant under the $O_{m-n}$ action.

Here, we need to introduce some more
 geometric quantities associated with the embedding, which does not appear 
in the case of the codimension-1 embeddings (see, e.g., Ref.~\cite{Eis26}).
Define the covariant vector field on $N$, parametrized by $(A,B)$, by
\begin{align*}
\mu_{ABa}=g_{\mu\nu}n_A{}^\mu D_a n_B{}^\nu+\Gamma_{\alpha\beta\gamma}
n_A{}^\alpha n_B{}^\beta D_a X^\gamma,
\end{align*}
which is essentially the projection of $n_A{}^\mu\nabla_\nu n_{B\mu}$ onto the
cotangent space on $N$, i.e. a part of the Cartan's connection coefficients 
appearing in the structure equations for the orthonormal frame in $M$.
Note that $n_A{}^\mu$ is regarded as a scalar function on $N$ in the above expression so that $D_a$ denotes just a partial derivative with respect to $\xi^a$.
These $\mu_{ABa}$'s are not independent but subject to
\begin{align*}
\mu_{ABa}=\mu_{[AB]a}.
\end{align*}
 It appears in the orthogonal decomposition of
$D_an_A{}^\mu$ as
\begin{align}
D_a n_A{}^\mu=-K_A{}^b{}_a D_bX^\mu-\mu_A{}^B{}_a n_B{}^\mu-\Gamma^\mu_{\nu\lambda}
(D_aX^\nu)n_A{}^\lambda
\label{Dn},
\end{align}
as readily confirmed (see, e.g. Refs.~cite{Eis26}),
where we have defined
\begin{align*}
  K_{Aab}=n_{A\mu}K^\mu{}_{ab}.
  \end{align*}

According to the small displacement $X^\mu\mapsto X^\mu+f^An_A{}^\mu$,
the induced metric on the membrane undergoes a variation $\delta G_{ab}$,
which in the first order of $f^A$'s is given by
\begin{align*}
\delta G_{ab}=-2K_{Aab}f^A.
\end{align*}
The variation of $\sqrt{|G|}D_aD^a X^\mu$ becomes
\begin{align}
\delta (\sqrt{|G|}D_a D^a X^\mu)
&=\sqrt{|G|}\biggl\{
n_A{}^\mu D_aD^a f^A
+\left[2\wtK_A{}^{ab}(D_bX^\mu)
+2(D^an_A{}^\mu)\right]D_af^A\nonumber\\
&+\left[2(D_a \wtK_A{}^{ab})(D_bX^\mu)
+2\wtK_A{}^{ab}(D_aD_b X^\mu)+(D_aD^an_A{}^\mu)\right ]
f^A
\biggr\}.
\label{1st}
\end{align}
Thus, we need the expression for the d'Alembertian of $n_A{}^\mu$,
which becomes
\begin{align}
D_aD^an_A{}^\mu&=
\left[-(D_aK_A{}^{ab})
+\mu_A{}^B{}_aK_B{}^{ab}\right]D_bX^\mu
-K_A{}^{ab}K^\mu{}_{ab}
\nonumber\\
&+2\Gamma^\mu_{\alpha\beta}(D_aX^\alpha)[
K_A{}^{ab}D_bX^\beta+\mu_A{}^{Ba}n_B{}^\beta]\nonumber\\
&+\left[-D_a\mu_A{}^{Ca}+\mu_A{}^B{}_a\mu_B{}^{Ca}
\right]n_C{}^\mu\nonumber\\
&+(-\Gamma^\mu_{\alpha\gamma,\beta}+\Gamma^\mu_{\gamma\delta}\Gamma^\delta_{\alpha\beta}
+\Gamma^\mu_{\beta\delta}\Gamma^\delta_{\alpha\gamma})G^{\alpha\beta}n_A{}^\gamma,
\label{DDn}
\end{align}
where
\begin{align*}
G^{\alpha\beta}=G^{ab}(D_aX^\alpha)D_bX^\beta
\end{align*}
is the  component of $G^{ab}$ written in terms of the spacetime coordinate
system.

Next, the first variation of the second terms of the Nambu-Goto equation
results in
\begin{align}
\delta (\sqrt{|G|}\Gamma^\mu_{\alpha\beta}G^{\alpha\beta})
&=\sqrt{|G|}\biggl\{
2\Gamma^\mu_{\alpha\beta}(D^aX^\alpha)n_A{}^\beta D_a f^A\nonumber\\
&+\Gamma^\mu_{\alpha\beta}\left[\wtK_A{}^{ab}(D_a X^\alpha)(D_bX^\beta)
-2\mu_A{}^{Ba}(D_aX^\alpha)n_B{}^\beta\right]f^A\nonumber\\
&+\left[
(\Gamma^\mu_{\alpha\beta,\gamma}-2\Gamma^\mu_{\alpha\delta}\Gamma^\delta_{\beta\gamma})
G^{\alpha\beta}n_A{}^\gamma
\right]f^A
\biggr\}.
\label{2nd}
\end{align}
Finally, from Eqs.~(\ref{1st}) and (\ref{2nd}),
using Eqs.~(\ref{DDX}), (\ref{Dn}), and (\ref{DDn}),
we obtain 
\begin{align}
\delta(\sqrt{|G|}(D_aD^aX^\mu+\Gamma^\mu_{\alpha\beta}G^{\alpha\beta})
&=\sqrt{|G|}( L f^C)n_C{}^\mu,
\end{align}
where 
\begin{align}
Lf^C&=D_aD^a f^C-2\mu_A{}^{Ca}D_af^A
+\biggl[
K_A{}^{ab}K^C{}_{ab}
-D_a\mu_A{}^{Ca}+\mu_A{}^B{}_a\mu_B{}^{Ca}\nonumber\\
&+R_{\alpha\beta}n_A{}^\alpha n^{C\beta}
-R_{\alpha\beta\gamma\delta}n_A{}^\alpha n_D{}^\beta n^{C\gamma}n^{D\delta}
\biggr]f^A,
\end{align}
which involves the spacetime Riemann and Ricci curvatures 
defined by
\begin{align*}
R^\mu{}_{\nu\lambda\rho}&=\Gamma^\mu_{\nu\rho,\lambda}-
\Gamma^\mu_{\nu\lambda,\rho}
+\Gamma^\mu_{\lambda\delta}\Gamma^\delta_{\nu\rho}
-\Gamma^\mu_{\rho\delta}\Gamma^\delta_{\nu\lambda},\\
R_{\nu\rho}&=R^\mu{}_{\nu\mu\rho}.
\end{align*}
Hence, the linear perturbation equation for the Nambu-Goto membranes are
governed by 
\begin{align*}
Lf^C=0.
\end{align*}

For consistency, let us confirm the covariance of $Lf^C=0$ under the
local $O_{m-n}$ transformation
\begin{align*}
n_A{}^\mu\mapsto O_A{}^Bn_B{}^\mu
\end{align*}
in terms of an orthogonal matrix field $O_A{}^B$ on $N$.
This is regarded as the gauge transformation in the principal
$O_{m-n}$ bundle over $N$.
The quantities $f^A$ and $K^A{}_{ab}$  transform like tensors as
\begin{align*}
(f^A,K^A{}_{ab})\mapsto O^A{}_B(f^B,K^B{}_{ab}),
\end{align*}
while  $\mu_A{}^B{}_a$ transforms like the principal $O_{m-n}$ connection as
\begin{align*}
\mu_{A}{}^B{}_a\mapsto 
O_A{}^CD_a O^B{}_C
+O_A{}^C\mu_C{}^D{}_a O^B{}_D,
\end{align*}
where the first term can be regarded as the pure gauge.
These guarantee the covariance of the perturbation equation:
\begin{align*}
Lf^C\mapsto O^C{}_D Lf^D.
\end{align*}

With this interpretation of $\mu_A{}^B{}_a$ as the connection on the
principal $O_{m-n}$-bundle, which is an $\frak{o}_{m-n}$-valued 1-form on $N$,
it turns out that the perturbation equation can be written more
compactly as
\begin{align}
L f^B=\mathscr{D}_a\mathscr{D}^a f^B
+\left(K_A{}^{ab}K^B{}_{ab}
+R_{\alpha\beta}n_A{}^\alpha n^{B\beta}
-R_{\alpha\beta\gamma\delta}n_A{}^\alpha n_C{}^\beta n^{B\gamma}n^{C\delta}
\right)f^A=0,
\label{p-eq}
\end{align}
where the covariant derivative
\begin{align*}
\mathscr{D}_af^B=D_a f^B-\mu_A{}^B{}_a f^A
\end{align*}
acting on  sections of the associated vector bundle is defined.
This expression for the perturbed Nambu-Goto equation 
is manifestly covariant under the $O_{m-n}$ gauge
transformation and the diffeomorphism on $N$.

\section{Stationary closed strings in 5-dimensional flat spacetimes}

Here, we consider stationary Nambu-Goto strings winding around a flat torus in the five-dimensional flat spacetime as a reference for a later section.

Starting with the line element
\begin{align*}
  g=-(dx^0)^2+(dx^1)^2+(dx^2)^2+(dx^3)^2+(dx^4)^2,
\end{align*}
we take a coordinate system $(t,r,\phi,\rho,\psi)$ determined by
\begin{align*}
  x^0=t,~~  x^1=r\cos\phi,~~
  x^2=r\sin\phi,~~
  x^3=\rho\cos\psi,~~
  x^4=\rho\sin\psi.
\end{align*}
Here and in what follows, we set the speed of light to unity.
Then, the line element takes the form
\begin{align*}
  g=-dt^2+dr^2+r^2d\phi^2+d\rho^2+\rho^2 d\psi^2
\end{align*}
with these coordinates.
It admits a simple solution to the Nambu-Goto equation
\begin{align*}
  t=\dfrac{pq}{\sqrt{p^2+q^2}}R\tau,~~
        r=\dfrac{q}{\sqrt{p^2+q^2}}R,~~
    \rho=\dfrac{p}{\sqrt{p^2+q^2}}R,~~
  \phi=p \sigma,~~
  \psi=q(\sigma+\tau),
\end{align*}
where $\tau\in\boldsymbol{R}$ and $\sigma\in \boldsymbol{R}/2\pi\boldsymbol{Z}$ are world sheet coordinates,
$p$ and $q$ are coprime integers,
and $R$ is a positive real number.
This describe a closed string winding around a torus at $(r,\rho)={\rm const}.$
with the winding number characterized by the coprime pair $(p,q)$,
and the closed string is stationarily scrolling on the torus.

We consider the linear perturbation of this solution.
Since the Nambu-Goto equation in flat space reduces to
a linear wave equation,
\begin{align*}
  D_a D^a X^\mu=0,
\end{align*}
its perturbation $\delta X^\mu$ is also subject to the same equation:
\begin{align*}
 D_a D^a\delta X^\mu=0.
\end{align*}
Its solution generally contains diffeomorphism on the world sheet,
which is unphysical.
On the other hand, Eq.~(\ref{p-eq}) describes only physical modes
contained in $\delta X^\mu$.

Now, we set 
\begin{align*}
  n_1&=\partial_r,\\
  n_2&=\partial_\rho,\\
  n_3&=\partial_t-\dfrac{\sqrt{p^2+q^2}}{pqR}(p\partial_\phi-q\partial_\psi),
\end{align*}
as an orthonormal frame $(n_1,n_2,n_3)$ for the normal space to the world sheet.
We need the following geometric quantities:
\begin{align*}
  G_{ab}&=\dfrac{p^2q^2R^2}{p^2+q^2}\left(
  \begin{array}{cc}
    0&1\\
    1&2
  \end{array}
  \right),\\
  %
  K^1{}_{ab}&=-\dfrac{p^2 q R}{\sqrt{p^2+q^2}}
  \left(\begin{array}{cc}0&0\\0&1  \end{array}  \right),~~
    K^2{}_{ab}=-\dfrac{pq^2 R}{\sqrt{p^2+q^2}}
    \left(\begin{array}{cc}1&1\\1&1    \end{array}\right),~~
    K^3{}_{ab}=0,   \\
  \mu_A{}^B{}_\tau&=\left(
  \begin{array}{ccc}
    0&0&0\\
    0&0&-q\\
    0&q&0
    \end{array}
  \right),~~
  \mu_A{}^B{}_\sigma=\left(
  \begin{array}{ccc}
    0&0&p\\
    0&0&-q\\
    -p&q&0
    \end{array}
  \right).  
\end{align*}
Then, assuming $f^A\propto e^{-i\omega\tau+ik\sigma}$ $(k\in\boldsymbol{Z})$,
the perturbation equation (\ref{p-eq}) reduces to
the algebraic equation
\begin{align*}
A\left(
  \begin{array}{c}
    f^1\\f^2\\f^3
  \end{array}
  \right)=0,~~~
  A=\left(
\begin{array}{ccc}
  \omega(\omega+k)&pq&-ip\omega\\
  pq&\omega(\omega+k)&-iq(\omega+k)\\
  ip\omega&iq(\omega+k)&\omega(\omega+k)
  \end{array}
  \right).
\end{align*}
The condition that this equation admits a nontrivial solution for
$f^A$'s is determined by
\begin{align*}
  \operatorname{det}A=\omega(\omega+k)(\omega+k+p)(\omega+k-p)(\omega+q)(\omega-q)=0,
\end{align*}
and hence, it is given by
\begin{align*}
  \omega=0,~~-k,~~-k\pm p,~~\pm q.
\end{align*}
All these modes show the stability of the Nambu-Goto strings in flat
background, as expected.

\section{Stationary closed strings in 5-dimensional black-hole space-times}
As a straightforward extension to the example given in the previous section,
we consider the stationary closed strings winding around a flat torus
in the five-dimensional Schwarzschild spacetime.
This turns out to be one of simplest cases allowing the analytic treatment of the perturbation
equation, which is one of reasons why
we describe it here in this paper.
Here, we show that this type of closed strings is generally unstable under the
small perturbation in the presence of the black holes.

The line element of the five-dimensional Schwarzschild spacetime is
given by
\begin{align*}
  g=-\left(1-\dfrac{r_0^2}{r^2}\right)dt^2
  +\left(1-\dfrac{r_0^2}{r^2}\right)^{-1}dr^2
  +r^2[d\theta^2+(\sin\theta)^2d\phi^2+(\cos\theta)^2d\psi^2],
\end{align*}
where $r_0>0$  corresponds to the Schwarzschild radius
of the event horizon,
$\theta\in (0,\pi/2)$, $\phi\in \boldsymbol{R}/2\pi\boldsymbol{Z}$, $\psi\in \boldsymbol{R}/2\pi\boldsymbol{Z}$ 
are the coordinates on the 3-sphere given by $t,r={\rm const}.$

This admits a Nambu-Goto string solution,
\begin{align}
  t=\dfrac{sr_0\tau}{\sqrt{2}},~~
  r=sr_0,~~
  \theta=\dfrac{\pi}{4},~~
  \phi=\sigma,~~
  \psi=\sigma+\tau,
\label{ng-sol}
\end{align}
where $\tau\in\boldsymbol{R}$ and $\sigma\in\boldsymbol{R}/2\pi\boldsymbol{Z}$ are world sheet coordinates and
$s>\sqrt{2}$ is a unique parameter of this solution
characterizing the distance between
the string and the black hole.
This describes a stationarily scrolling closed string
winding around a flat torus embedded in the Schwarzschild spacetime,
with the winding number $(p,q)=(1,1)$.
We note that this Nambu-Goto string is just a special case of the solutions
considered by Igata and Ishihara~\cite{II10a,II10b}.

We choose
\begin{align*}
  n_1&=\dfrac{\sqrt{s^2-1}}{s}\partial_r,\\
  n_2&=\dfrac{1}{sr_0}\partial_\theta,\\
  n_3&=\dfrac{s^2}{\sqrt{(s^2-2)(s^2-1)}}\partial_t
    -\dfrac{\sqrt{2(s^2-1)}}{sr_0\sqrt{s^2-2}}(\partial_\phi-\partial_\psi),
\end{align*}
as a orthonormal frame $(n_1,n_2,n_3)$ for the normal space of the
world sheet.

The geometric quantities required to compute the perturbation equation are
\begin{align*}
  G_{ab}&=\dfrac{r_0^2}{2}\left(\begin{array}{cc}1&s^2\\s^2&2s^2
  \end{array}\right),\\
  K^1{}_{ab}&=-\dfrac{\sqrt{s^2-1}r_0}{2s^2}\left(
  \begin{array}{cc}s^2-1&s^2\\s^2&2s^2
  \end{array}\right),~~
  K^2{}_{ab}=\dfrac{s r_0}{2}\left(
  \begin{array}{cc}1&1\\1&0
  \end{array}\right),~~
      K^3{}_{ab}=0,\\
  \mu_A{}^B{}_\tau&=\left(
  \begin{array}{ccc}
    0&0&-\dfrac{\sqrt{s^2-2}}{\sqrt{2}s}\\
    0&0&\dfrac{\sqrt{s^2-1}}{\sqrt{2(s^2-2)}}\\
    \dfrac{\sqrt{s^2-2}}{\sqrt{2}s}&-\dfrac{\sqrt{s^2-1}}{\sqrt{2(s^2-2)}}&0
  \end{array}
  \right),\\
  \mu_A{}^B{}_\sigma&=\left(
  \begin{array}{ccc}
    0&0&0\\
    0&0&\dfrac{\sqrt{2(s^2-1)}}{\sqrt{s^2-2}}\\
    0&-\dfrac{\sqrt{2(s^2-1)}}{\sqrt{s^2-2}}&0
  \end{array}
  \right),\\
&  R_{\alpha\beta}n_A{}^\alpha n^{B\beta}=0,\\
 & R_{\alpha\beta\gamma\delta}n_A{}^\alpha n_C{}^\beta n^{B\gamma}n^{C\delta}
  =\dfrac{2}{s^4(s^2-2)r_0^2}\left(
  \begin{array}{ccc}
    2-3s^2&0&0\\
    0&s^2&0\\
    0&0&-s^2
  \end{array}\right).
\end{align*}
Assuming $f^A\propto e^{-i\omega \tau+ik\sigma}$ $(k\in\boldsymbol{Z})$,
Eq.~(\ref{p-eq}) becomes
\begin{align*}
&  A\left(\begin{array}{c}f^1\\f^2\\f^3
  \end{array}\right)=0,\\
&  A=\left(\begin{array}{ccc}
    2s^2\omega^2+2s^2 k\omega+k^2+2s^2-2&0&-is\sqrt{2(s^2-2)}(2\omega+k)\\
    0&2s^2\omega^2+2s^2 k\omega+k^2-2s^2&ik\sqrt{2(s^2-2)(s^2-1)}\\
    is\sqrt{2(s^2-2)}(2\omega+k)&-ik\sqrt{2(s^2-2)(s^2-1)}&
    2s^2\omega^2+2s^2k\omega+k^2
  \end{array}
  \right).
\end{align*}
Then, $f^A$'s have nontrivial solutions when
$\omega$ solves the polynomial equation
\begin{align*}
  \operatorname{det}A&=
  8s^6\omega^6+24s^6 k\omega^5
  +[12s^4(2s^2+1)k^2 -8s^4(2s^2-3)]\omega^4\\
&  +[8s^4(s^2+3)k^3-16s^4(2s^2-3)k]\omega^3\\
&  +[6s^2(2s^2+1)k^4-12s^4(2s^2-3)k^2+8s^4(s^2-3)]\omega^2\\
 & +[6s^2 k^5-4s^4(2s^2-3)k^3+8s^4(s^2-3)k]\omega\\
&  +[k^6-(4s^4-10s^2+6)k^4+(4s^4-16s^2+8)k^2]=0.
\end{align*}
Substituting $\omega=\sqrt{x}-k/2$, this reduces to
\begin{align}
\nonumber p(x)&=  8s^6 x^3+[6s^4(2-s^2)k^2+8s^4(3-2s^2)]x^2\\
\nonumber&  +\left[\dfrac{3}{2}s^2(s^2-2)^2k^4+8s^4(s^2-3)\right]x\\
  &  +
    \dfrac{1}{8}(2-s^2)^3k^6+\dfrac{1}{2}(s^2-2)^2(2s^2-3)k^4
    -2(s^2-2)^2(s^2-1)k^2
    =0.\label{px}
\end{align}
This is a cubic polynomial equation; hence, it is solvable via 
 Cardano's method.

Therefore, the present Nambu-Goto string is unstable if and only if
the polynomial $p(x)$ has two  complex-conjugate roots or a negative root,
which depends on the parameters $s$ and $k$.

It is easily seen that $p(x)$ has a negative root, when the string
is sufficiently close to the $r=\sqrt{2}r_0$ surface.
Setting $s^2=2+\epsilon$ $(\epsilon>0)$,
Eq.~(\ref{px}) becomes
\begin{align*}
  p(x)=8x[(8+12\epsilon)x^2-(4+3k^2\epsilon+12\epsilon)x-4]+O(\epsilon^2)=0,
\end{align*}
hence $p(x)$ has roots
\begin{align*}
  x=0,~~1+\dfrac{k^2}{4}\epsilon,~~-\dfrac{1}{2}+\dfrac{k^2+6}{8}\epsilon,
\end{align*}
up to first order in $\epsilon$.
The third root corresponds to the unstable mode behaving like
\begin{align*}
f^A\propto \exp\left[\left(\dfrac{1}{\sqrt{2}}-\dfrac{6+k^2}{8\sqrt{2}}\epsilon+i\dfrac{k}{2}\right)\tau\right]e^{ik\sigma}.
\end{align*}

For $s\gg 1$, since the geometry around the string approaches the flat
spacetime, one might expect that such strings are always stable.
We, however, show that it is not the case.

To see this, consider the expansion 
\begin{align*}
\nonumber p(x)&=  8s^6
\left(x-\dfrac{k^2}{4}\right)\left[x-\dfrac{(k-2)^2}{4}\right]
\left[x-\dfrac{(k+2)^2}{4}\right]
+O(s^{4})
    =0.
\end{align*}
From this expression, the approximate roots for $p(x)$ can be read off
from the leading $O(s^6)$ term. It can be seen that for $|k|\ge 2$ they are given by
three distinct positive roots $k^2/4$, $(k\pm 2)^2/4$, which are consistent with
the results in the previous section. Although the exact roots might slightly differ
from these under small corrections,  these would still give positive roots, showing the stability
of strings under these modes. 
The cases $k=0,\pm 1$ should be considered separately,
when approximate roots include multiple one, since it possibly
becomes nonreal roots under small corrections.

The expression for small corrections from $O(s^{4})$ terms is readily
obtained thanks to  Cardano's formula. For $k=\pm 1$,
we can see that $p(x)$ has two complex-conjugate roots,
\begin{align*}
x=\dfrac{1}{4}-\dfrac{3}{8}s^{-2}\pm i\dfrac{\sqrt{15}}{8}s^{-2}+O(s^{-4})
\end{align*}
which correspond to the unstable modes
\begin{align*}
f^A&\propto \exp\left\{
\left[
\dfrac{\sqrt{15}+3i}{8}s^{-2}+O(s^{-4})
\right]
\tau\right\}
e^{\pm i\sigma},\\
f^A&\propto \exp\left\{
\left[i+
\dfrac{\sqrt{15}-3i}{8}s^{-2}+O(s^{-4})
\right]
\tau\right\}
e^{\pm i\sigma},
\end{align*}
These instabilities, however, are exposed after a relatively long latent period 
$\tau\sim s^2$, so  strings might be possibly stabilized taking into account 
dissipative effects, such as the emission of Nambu-Goldstone bosons, 
which are not considered in the present analysis of
test strings. We also find out that the uniform $k=0$ modes do not show such
instabilities, where $p(x)$ has exact roots $0$, $1$, $1-3s^{-2}$.

We finally show that $k=\pm 1$ modes are always unstable.
\begin{thm}
  The Nambu-Goto string solutions given by  Eqs.~(\ref{ng-sol})
  are unstable under the linear perturbation.
\end{thm}

\noindent{\em Proof.}
The statement of  Theorem 1 is proven by showing that the cubic polynomial $p(x)$ has two complex-conjugates roots
and 
otherwise  has at least one negative roots, when $k=\pm 1$.

We can easily see that  $p(x)$ always has
the local maximum at
\begin{align*}
x=x_1:=\dfrac{11}{12}-\dfrac{3}{2s^2}-\dfrac{\sqrt{16s^4-54s^2+72}}{6s^2},
\end{align*}
when $k=\pm 1$.

This is an increasing function of $s$ for $s>\sqrt{2}$,
so  it can be shown that $x_1$ is negative for $\sqrt{2}<s<s_1$,
where $s_1$ is given by
\begin{align*}
s_1=\sqrt{\dfrac{30+4\sqrt{42}}{19}}\approx 1.7.
\end{align*}

On the other hand, the corresponding local maximum value of $p(x)$ is given by
\begin{align*}
p(x_1)=-\dfrac{128}{27}s^6+24s^4-56s^2+48
+\left(\dfrac{32}{27}s^4-4s^2+\dfrac{16}{3}\right)\sqrt{16s^4-54s^2+72}.
\end{align*}
As a function of $s$, the zeros of $p(x_1)$ can be determined
by solving a quartic equation for $s^2$.
Then, we find that $p(x_1)$ has
only one zero for $s>\sqrt{2}$ at
\begin{align*}
s=s_2=\sqrt{2+\left(\dfrac{2}{5}\right)^{2/3}(5+3\sqrt{5})^{1/3}
-2\left(\dfrac{2}{5}\right)^{1/3}(5+3\sqrt{5})^{-1/3}}\approx 1.6.
\end{align*}
It turns out that
the local maximum value $p(x_1)$ is negative for $s>s_2$, 
In particular, $s_2$ is less than $s_1$.

Then, it can be concluded that
$p(x)$ has two  complex-conjugate roots for $s>s_2$ and
that $p(x)$ has at least one negative root (in fact, it always has exactly
two negative roots) for $\sqrt{2}<s\le s_2$, when $k=\pm 1$.

Therefore, the Nambu-Goto strings given by Eqs.~(\ref{ng-sol})
are always unstable.\\
{\flushright$\Box$\par\medskip}

\section{Conclusions}
We have studied the stability of stationary closed strings
winding around a flat torus embedded in the five-dimensional Schwarzschild
spacetime.
The Nambu-Goto strings belonging to this class are characterized by a real
parameter $s>\sqrt{2}$, with which  the location of the closed string is
written as $r=sr_0$, where $r_0$ denotes the Schwarzschild radius.
We have shown that the perturbation modes are calculable
only with algebraic manipulations.
We have proven that all the solutions belonging to this class
are unstable under linear perturbations.

\end{document}